\documentclass[usenatbib]{mn2e}

\usepackage{epsfig,times,amsmath,supertabular,colortbl,amsfonts,amssymb,multirow,array,fleqn,longtable}

\title[The SZ effect due to hyper-starburst winds]{The Sunyaev-Zel'dovich effect due to hyper-starburst galaxy winds}

\author[B.\ Rowe \& J.\ Silk]{Barnaby~Rowe,$^{1,2}$\thanks{E-mail: barnaby.t.rowe@jpl.nasa.gov}
Joseph~Silk$^{3}$ \\ 
$^1$Jet Propulsion Laboratory, California Institute of Technology, 4800 Oak Grove Drive, Pasadena, California 91109, USA \\
$^2$Institut d'Astrophysique de Paris, UMR7095 CNRS, Universit\'{e}
Pierre et Marie Curie -- Paris 6, 98 bis, Boulevard Arago, 75014
Paris, France \\
$^3$Department of Physics, Oxford University, Denys Wilkinson Building, Keble Road, Oxford, OX1 4LN, UK}

\begin{document}
\maketitle

\newcommand{\expect}[1]{\left\langle #1 \right\rangle} 
\newcommand{\dif}{\mbox{$\mathrm{d}$}}
\newcommand{\mobs}{\mbox{\scriptsize$\mathrm{obs}$}}
\newcommand{\mcrit}{\mbox{\scriptsize$\mathrm{crit}$}}
\newcommand{\mres}{\mbox{\scriptsize$\mathrm{res}$}}
\newcommand{\mQs}{\mbox{\tiny$\mathrm{Q}$}}
\newcommand{\mw}{\mbox{\scriptsize$\mathrm{w}$}}
\newcommand{\mb}{\mbox{\scriptsize$\mathrm{b}$}}
\newcommand{\thetab}{\mbox{\boldmath$\theta$}}
\newcommand{\xb}{\mbox{\boldmath$x$}}
\newcommand{\rb}{\mbox{\boldmath$r$}}
\newcommand{\varthetab}{\mbox{\boldmath$\vartheta$}}
\newcommand{\phib}{\mbox{\boldmath$\phi$}}
\newcommand{\varphib}{\mbox{\boldmath$\varphi$}}
\newcommand{\ms}{\mbox{\scriptsize$\mathrm{s}$}}
\newcommand{\mi}{\mbox{$\mathrm{i}$}}

\newcommand{\nat}{Nat}
\newcommand{\mnras}{MNRAS}
 \newcommand{\apj}{ApJ}
\newcommand{\apjl}{ApJL}
\newcommand{\apjs}{ApJS}
\newcommand{\physrep}{Phys.~Rep.}
\newcommand{\aap}{A\&A}
\newcommand{\aaps}{A\&AS}
\newcommand{\aj}{AJ}
\newcommand{\araa}{ARA\&A}
\newcommand{\apss}{Ap\&SS}
\newcommand{\pasp}{PASP}

\begin{abstract}
  We construct a simple, spherical blastwave model to estimate the pressure   structure of the intergalactic medium surrounding hyper-starburst galaxies, and argue that the effects of interaction with star-forming galaxy winds may   be approximated at early times by an adiabatically expanding, self-similar   `bubble' as described by \citet{weaveretal77} and \citet{ostrikermckee88}. This model is used to make observational predictions for the thermal Sunyaev-Zel'dovich effect in the shocked bubble plasma.  Radiative cooling losses are explored, and it is found that bremsstrahlung will limit the epoch of adiabatic expansion to $10^7$--$10^8$ years: comparable to total hyper-starburst lifetimes.  Prospects for making a first Sunyaev-Zel'dovich detection of galaxy wind bubbles using the Atacama Large Millimeter Array are examined for a number of active hyper-starburst sources in the literature.
\end{abstract}


\section{Introduction}
After the epoch of reionization, free electrons in the intergalactic medium (IGM) interact with photons in the cosmic microwave background radiation (CMB) via Thompson scattering.  Known as the Sunyaev-Zel'dovich (SZ) effect (e.g.\ \citealp{sunyaevzeldovich70,sunyaevzeldovich72}; see also \citealp{carlstrometal02} for a review), the result is a characteristic spectral distortion of the CMB.  The effect can be split into two contributions: that due to the thermal motion of free electrons which causes an overall increase in the apparent temperature of CMB photons (called the thermal SZ or tSZ effect), and that due bulk kinetic motion (the kSZ effect; see \citealp{carlstrometal02}).

The tSZ effect, in particular, shows promise as a means of observing the hot plasma surrounding galaxy clusters and other peaks in the matter power spectrum, and recent years have seen a steady growth in both theoretical predictions (e.g.\ \citealp{moodleyetal09}; \citealp*{scannapiecoetal08}; \citealp{roncarellietal07}) and actual measurements for massive galaxy clusters (e.g.\ \citealp{halversonetal09,basuetal10,nordetal09}). It is interesting to consider what other astrophysical objects may be observable in the near future with the tSZ effect (e.g.\ \citealp{yamadaetal10}).

In recent years an increasing number of high redshift `hyper-starbursts' (which we define as objects with an estimated star formation rate of $\dot{M}_{*} \gtrsim 10^3$ M$_{\odot}$yr$^{-1}$) have been reported in the literature, often but not always identified with submillimetre galaxies (SMGs), and are believed to represent a significant fraction of the total star formation at these epochs (e.g.\ \citealp{blainetal02,solomonvandenbout05,wangetal08,caseyetal09,martinezetal09,riechersetal09,waggetal09}).
Other recent results suggest that the galaxy wind outflows frequently observed in star-forming galaxy spectra (see, e.g., \citealp*{veilleuxetal05} for a review) are ubiquitous wherever there is star formation activity to drive them \citep{grimesetal09,rubinetal10,weineretal09,krugetal10}, with wind masses and velocities increasing with $\dot{M}_{*}$. The winds associated with hyper-starbursts at high redshift will be extremely energetic, leading to violent interaction with the plasma of the surrounding intergalactic medium (IGM).

The following question naturally arises: can we detect the tSZ due to the wind-IGM interaction for these events?  A similar question has been posed for active galaxies (e.g.\ \citealp*{yamadaetal99}; \citealp{plataniaetal02,chatterjeekosowsky07,scannapiecoetal08,chatterjeeetal08,yamadaetal10}) and for the combined signal due to star-forming winds (e.g.\ \citealp*{majumdaretal01,babichloeb07}), but individual hyper-starburst objects now also raise interesting possibilities. 
The temperature increase of CMB photons due to the tSZ effect is redshift independent (e.g.\ \citealp{carlstrometal02}), which potentially opens a new observational window on extreme star formation at high redshift. 

We examine this question by constructing a simplified adiabatic model of the wind-IGM interaction (Section \ref{sect:bubble}), and then calculate the tSZ effect according to such a model for a variety of possible hyper-starburst targets in the literature (Sections \ref{sect:sz} \& \ref{sect:targets}) along with observational prospects for the upcoming Atacama Large Millimeter Array (ALMA: see \citealp*{brownetal04}) facility.  In Sections \ref{sect:cooling} we test the adiabatic expansion assumption by calculating radiative cooling timescales for the hot plasma, and end in Section \ref{sect:conc} with a discussion and conclusions.

\section{Galaxy wind injection bubble model}\label{sect:bubble}
As demonstrated by \citet*{samuietal08}, who present a detailed semi-analytic analysis of galactic outflows, the galaxy wind-IGM interaction can be approximately described at early times as a spherically expanding, self-similar blastwave or `bubble'.  The model used to describe this bubble is based on that presented in \citet{weaveretal77}, which describes the adiabatic interaction of strong stellar winds with the interstellar medium. This is related to the original Sedov-Taylor blastwave solution (see, e.g., \citealp{landaulifshitz59,ostrikermckee88}) but describes the shock evolution under \emph{continuous} rather than instantaneous energy injection, making it being better suited to cases with ongoing starburst activity.  The \citet{weaveretal77} model retains self-similarity by not introducing a characteristic timescale to the stellar wind bubble, and it is this property that allows it to be applied equally well to the galaxy wind-IGM interaction despite the greater energies of this regime.  

At time $t=0$ star formation processes initiate the expulsion of a galaxy wind, which we approximate as having constant mass outflow rate $\dot{M}_{\mw}$ leading to a mechanical energy injection rate given by
\begin{equation}
L_{\mw} = \dot{M}_{\mw} v^2_{\mw} / 2
\end{equation}
where $v_{\mw}$ is the net wind outflow velocity. All quantities will be expressed in relation to a fiducial value $v_{\mw} = 10^3$ kms$^{-1}$, motivated by observations of intermediate-to-high redshift winds for objects with extreme star formation rates \citep{rubinetal10, weineretal09,capaketal08,blandhawthornetal07,adelbergeretal03}.
Following \citet{aguirreetal01}, \citet{springelhernquist03}, and \citet{dallavecchiaschaye08}, we relate the wind mass outflow $\dot{M}_{\mw}$ to the star formation rate $\dot{M}_*$ using the dimensionless wind mass loading parameter $\eta$ (adopting a fiducial value of 1 in all calculations), so that $\dot{M}_{\mw} = \eta \dot{M}_* $. 
The total energy injected by the wind into the system is thus
\begin{equation}
E= L_{\mw} t= \eta \dot{M}_* v^2_{\mw} t / 2.
\end{equation}
The mechanical luminosity $L_{\mw}$ can be related to the total energy available from core collapse supernovae via the wind energy fraction
\begin{equation}
f_{\mw} = \eta v_{\mw}^2 / (2 \epsilon_{\textrm{SN}}) ,
\end{equation}
where $\epsilon_{\textrm{SN}}$ is the kinetic energy injected per M$_{\odot}$ of stars formed (e.g.\ \citealp{dallavecchiaschaye08}). Assuming a \citet{chabrier03} initial mass function, these authors adopt a value $\epsilon_{\textrm{SN}} \simeq 1.8 \times 10^{42} \textrm{erg} \textrm{M}^{-1}_{\odot}$: for our fiducial values of $v_{\mw}$ and $\eta$ this implies a wind energy fraction of $f_{\mw} \simeq 0.55$. This is slightly larger than the value of $0.4$ adopted by \citet{dallavecchiaschaye08} for more moderately star-forming systems. It should be cautioned that the exact value of $f_{\mw}$ appropriate for supernovae in high-redshift hyper-starburst regions is unclear, and so all results will be presented with scaling relations for the parameters $v_{\mw}$ and $\eta$.

The interaction of the wind with the IGM can be split into four physically distinct zones at increasing radial distance $r$ from the bubble centre: a) the free-streaming wind immediately on exit from the hyper-starburst region; b) a region of shocked galaxy wind; c) a shell of shocked IGM plasma; d) the ambient IGM surrounding the hyper-starburst host galaxy
(\citealp{weaveretal77,samuietal08}).
We model the unshocked IGM in zone d) as ionized hydrogen of homogeneous density $\rho_d$. In order to estimate a value for this density in a hyper-starburst galaxy, we make use of measurements for the clustering bias $b_{\mQs}$ of high-redshift SDSS-DR5 quasars given in \citet{shenetal09}: large molecular gas reservoirs and hyper-starburst events are often associated with quasi-stellar objects (QSOs), radio galaxies, or other indications of active galactic nuclei (e.g.\ \citealp{solomonvandenbout05}; see also Section \ref{sect:targets}, Table \ref{tab:targets}).  
These authors find a bias of $b_{\mQs} = 12.96 \pm 2.09$ for the sample of 1788 objects in their highest redshift bin: $3.5 < z < 5.0$ (median $z=3.84$), and we thus choose a fiducial modelling value of $b_{\mQs} = 13$.  We also assume that the IGM in the region immediately surrounding the starburst is biased in the same way, giving
\begin{equation}
\rho_d \simeq \rho_{\mcrit} \Omega_{\mb} (1 + z)^3 (1 + b_{\mQs} \delta),
\end{equation}
where $\rho_{\mcrit}$ and $\Omega_{\mb}$ are the current-epoch critical density and baryon density parameter respectively, and $\delta$ is the dimensionless density perturbation.  We approximate this perturbation in the star formation environment as $\delta \simeq 180$, the mean overall matter density perturbation for a collapsed halo.

The outer radius of an adiabatically expanding (non-radiative), spherical blastwave under constant energy injection is given by
\begin{equation}
R_2 (t) = \beta \left( L_{\mw} t^3 / \rho_d \right)^{\frac{1}{5}} \label{eq:rdim},
\end{equation}
where $\beta = 0.8828$ is a constant estimated via numerical calculation \citep{ostrikermckee88}.  Equation \eqref{eq:rdim} is the only dimensionally correct combination of the system variables in this time and distance scale-free problem, and its power law form is therefore required; this self-similar evolution will cease only after radiative losses become significant (see Section \ref{sect:cooling}).  Using a flat, WMAP 5-year best-fitting $\Lambda$CDM cosmology \citep{komatsuetal09} and the value $\Omega_{\mb} = 0.044 \pm 0.01$ \citep{sanchezetal09}, we may rewrite equation \eqref{eq:rdim} in terms of fiducial values for a hyper-starburst shockwave system as follows:
 \begin{eqnarray}
R_2(t) & = & \beta \left[\frac{\eta \dot{M}_* v^2_{\mw} t^3}
{2 \Omega_{\mb} \rho_{\mcrit} (1 + z )^3(1 + b_{\mQs} \delta)} \right]^{\frac{1}{5}} \\
 & = & 28.86 \: {\eta}^{\frac{1}{5}} 
\left( \frac{\dot{M}_*}{10^3 \, \textrm{M$_{\odot}$yr}^{-1}} \right)^{\frac{1}{5}}
\left( \frac{v_{\mw}}{10^3 \: \textrm{kms}^{-1}} \right)^{\frac{2}{5}}
 \nonumber \\
& \times & \left( \frac{t}{10^7 \, \textrm{yr}} \right)^{\frac{3}{5}}
\left( 1 + z \right)^{-\frac{3}{5}} \textrm{kpc} \label{eq:rt}.
\end{eqnarray}
It can be seen that bubble radius will quickly extend significantly beyond the star forming regions into the surrounding IGM.

The pressure and density structure within the bubble is also approximately self-similar \citep{weaveretal77,samuietal08}, and can therefore be described in terms of the dimensionless radial parameter $\xi = r/R_2(t)$. At the very edge of the shock ($\xi=1$), the gas quantities are determined by the Rankine-Hugoniot conditions for a strong shock boundary:
\begin{equation}\label{eq:rankine}
\rho_2 = \rho_d (\gamma + 1) / (\gamma - 1),  \: \: \: \: \: \: \: \:
P_2 = 2  \rho_d \dot{R}^2_2 / (\gamma + 1)  ,
\end{equation}
where $\dot{R}_2 \equiv \dif R_2/ \dif t$, and for the monatomic gas we have adiabatic index $\gamma = 5/3$ (see \citealp{landaulifshitz59}).  Using this value gives an edge-of-shock density of $\rho_2 = 4 \rho_d$ and pressure $P_2 = 3 \rho_d \dot{R}^2_2 / 4$.  Within the outer shock radius, \citet{weaveretal77} find that the density $\rho(\xi)$ decreases inwards and drops suddenly to zero at the contact surface $R_c = 0.86 R_2$, the interface between the shocked IGM in zone c) and the shocked wind in zone b).  The pressure at this surface is $P(\xi = 0.86) = 0.59 \rho_d \dot{R}^2_2$, and these scalings hold while the expansion remains adiabatic.

\begin{figure}
\begin{center}
\psfig{figure=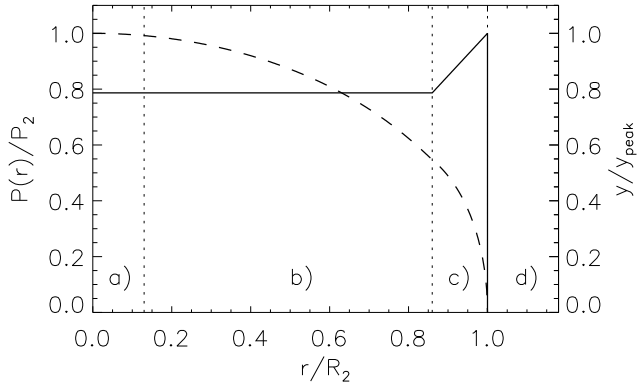,angle=0,width=8.6cm}
\parbox{8.4cm}{\caption{Pressure (solid line) and Compton $y$ parameter (dashed line) in the model as a function of distance from the bubble centre, with labels for the four distinct physical zones a)--d). \label{fig:bubble}}}
\end{center}
\end{figure}

For a perfect gas the tSZ effect may be estimated by integrating the pressure along a line-of-sight path through the bubble; a model for the pressure structure throughout must therefore be adopted.  An exact model may be calculated numerically \citep{weaveretal77,dokuchaev02} but ultimately the accuracy of such a model will rely on the physical conditions matching those described above (e.g.\ uniform ambient IGM density $\rho_d$ or constant input star-formation power $L_{\mw}$), which is unlikely.  Instead, for the purposes of estimating an approximate observability of the tSZ effect we adopt a simple model that displays the correct behaviour at the boundaries and makes simplifying interpolations between them:
\begin{equation}\label{eq:pfunc}
P(\xi)  = \left\{ \begin{array}{lc} P_b & 0 \le \xi < 0.86  \\
                         P_b + \left(\frac{P_2 - P_b}{0.14} \right) (\xi - 0.86) ~ ~ ~ & 0.86 < \xi \le 1 \end{array}
\right.
\end{equation}
where $P_b = 0.59 \rho_d \dot{R}^2_2$.  The model, plotted in Figure \ref{fig:bubble}, is isobaric for $r < R_c$: this behaviour is argued from simple principles by \citet{weaveretal77} and can also be seen in the complex analytic solutions provided by \citet{dokuchaev02}.  This pressure model can then be used to estimate the tSZ effect for a hyper-starburst wind bubble.

\section{Thermal SZ effect in the bubble}\label{sect:sz}

The tSZ effect describes the spectral variation of the specific intensity of the CMB photons caused by their passing through a non-relativistic, Maxwell-distributed electron gas \citep{carlstrometal02}.  It takes the form
\begin{equation}
\frac{\Delta I_{\nu}}{I_{\nu}} = \frac{x \exp{x}}{\exp{x} - 1} \left( \frac{x}{\tanh{x/2}} -4 \right) y
\end{equation}
with $x = h \nu /k_{\textrm{B}} T_{\textrm{CMB}}$ (using $T_{\textrm{CMB}} = 2.725$ K at $z=0$), and where the Compton $y$ parameter is defined via the path integral
\begin{equation}\label{eq:yT}
y = \int \frac{k_{\textrm{B}} T_e (l)}{m_e c^2} \sigma_{\textrm{T}} n_e(l) \dif l
\end{equation}
along the observer line-of-sight.  We do not include a relativistic correction to the tSZ effect (e.g.\ \citealp*{challinorlasenby98,itohetal98}) for the hot shockwave electrons (see Section \ref{sect:targets}).

For a perfect gas such as the extremely rarefied IGM, equation \eqref{eq:yT} becomes
\begin{equation}\label{eq:ypressure}
y = \int \frac{k_{\textrm{B}} P(l)}{m_e c^2} \sigma_{\textrm{T}} \dif l .
\end{equation}
Using the fiducial values described, and the pressure model described by equation \eqref{eq:pfunc}, this gives a peak value of
\begin{eqnarray}
y_{\textrm{peak}} & = & 2.456 \times 10^{-6}
\: {\eta}^{\frac{3}{5}} \left( \frac{\dot{M}_*}{10^3 \, \textrm{M$_{\odot}$yr}^{-1}} \right)^{\frac{3}{5}} \left( \frac{v_{\mw}}{10^3 \: \textrm{kms}^{-1}} \right)^{\frac{6}{5}}
 \nonumber \\
& \times & 
\left( \frac{t}{10^7 \, \textrm{yr}} \right)^{-\frac{1}{5}}
\left( 1 + z \right)^{\frac{6}{5}} \label{eq:ypeak}
\end{eqnarray}
along the line-of-sight through the centre of the bubble. Note that, due to the dependence of $P$ upon $\rho_d$, the strength of the tSZ in this model increases with redshift (although the bremsstrahlung cooling rate will also: see Section \ref{sect:cooling}).  The $y$ at any distance from the bubble centre can be calculated using the expression for pressure given in equation \eqref{eq:pfunc}, and is shown in Figure \ref{fig:bubble}.

\section{Potential hyper-starburst targets}\label{sect:targets}
\begin{table*}
\caption{Model predictions for a selection of reported hyper-starbursts, based on published redshifts and star formation rate estimates (see Sections \ref{sect:targets} \& \ref{sect:cooling}).}
\label{tab:targets}
\begin{center}
\begin{tabular}{lcccccccccc}
\hline
Galaxy target $^a$ & Redshift & Type & SFR & $t$ & $R_2$ & $\theta_2$ & $y_{\textrm{peak}}$ & $SN^{\textrm{peak}}_{24}$ & $\tau_{\textrm{IC}}$ & $\tau_{\textrm{B}}$ \\
 ~ & $z$ & ~& M$_{\odot} \textrm{yr}^{-1}$ & $10^7$yr &  kpc & arcsec & $10^{-5}$ & & $10^7$yr & $10^7$yr \\ 
\hline
RG J123649.66+620738.0$^b$ & 2.32 & Radio & 3800$^c$ & 1$^{\ddagger}$ & 18.4 & 2.21 & 2.32 & 1.19 &  971 & 36.0 \\
SMM J222174+0015$^d$ & 3.10 & SMG & 1800$^e$ & 1.7$^e$ & 19.1 & 2.47 & 1.71 & 0.88 & 410 & 11.6 \\
4C41.17R/B$^f$ & 3.80 & Radio & 3000$^e$ & 1.1$^e$ & 14.9  & 2.06 & 3.06 & 1.56 & 219 & 8.73 \\
SMM J154137+6630.5$^g$ & 3.93$^{\dagger}$ & SMG & $> 10 \, 000^g$ & 1$^{\ddagger}$ & 17.6 & 2.46 & 6.63 & 3.40 & 200 & 10.5 \\ 
COSMOS J100054+023436$^{h,i}$ & 4.55 & SMG & $2900^h$ & $0.7^h$ & 10.3 & 1.54 & 3.91 & 1.95 & 124 & 6.14 \\
BR 1202-0725$^j$ & 4.69 & QSO & 9000$^e$ & 0.9$^e$ & 14.8 & 2.27 & 7.55 & 3.97 & 112 & 6.37 \\
LESS J033229.4-275619$^k$ & 4.76 & SMG & $\sim 1000^k$ & $4^k$ & 23.2 & 3.53 & 1.52 & 0.82 & 106 & 2.16 \\ 
SDSS J092721.82+200123.7$^l$ & 5.77 & QSO & $3200$$^{m}$ & 0.5$^m$ & 7.63 & 1.28 & 5.63 & 2.42 & 56.3 & 3.50 \\
SDSS J114826.64+525150.3$^n$ & 6.42 & QSO & $3300^{o}$ & 1$^p$ & 11.1 & 1.97 & 5.67 & 2.89 & 38.8 & 1.93 \\
\hline 
\end{tabular}
\end{center}
\parbox{17.6cm}{{\small
$^a$(References for CO/FIR-emission detection or spectral confirmation); 
$^b$\citet{chapmanetal04};
$^c$\citet{caseyetal09};
$^d$\citet{nerietal03};
$^e$\citet{solomonvandenbout05};
$^f$\citet{debreucketal05}; $^g$\citet{waggetal09}; 
$^h$\citet{capaketal08}; $^i$\citet{schinnereretal08};
$^j$\citet{omontetal96}; $^k$\citet{coppinetal09}; 
$^l$\citet{carillietal07}; $^m$\citet{wangetal08}; $^n$\citet{bertoldietal03}; $^o$\citet{riechersetal09}; $^p$\citet{walteretal03}. \\
$^{\dagger}$Photometric redshift estimate. $^{\ddagger}$Unknown, fiducial value used (see equations \ref{eq:rt} \& \ref{eq:ypeak} for scalings).
}}
\end{table*}

In Table \ref{tab:targets} we present a limited selection of hyper-starburst objects from the literature that may make suitable targets for tSZ observation, along with predictions for observable properties based on the simple bubble model. In all these calculations we use the published estimate of the SFR, take $v_{\mw} = 10^3$ kms$^{-1}$ and $\eta=1$ (corresponding to $f_{\mw} \simeq 0.55$), and employ a flat, WMAP 5-year cosmology as in Section \ref{sect:bubble}.  Where an estimate of the age $t$ of the hyper-starburst exists (e.g.\ from SED fits) that value is chosen; otherwise, the depletion timescale for the CO-detected molecular gas reservoir is used as a proxy (since on average a hyper-starburst will be observed midway through this depletion).  The observational angular radius $\theta_2$ corresponding to $R_2$ is given for each bubble model, and the central line-of-sight Compton $y_{\textrm{peak}}$ is tabulated. In all cases this is seen to be significantly greater than the cosmological expectation value of $\expect{y} = (1.19 \pm 0.32) \times 10^{-6}$ calculated by \citet{roncarellietal07}.

Using the ALMA sensitivity calculator (made available on the European Southern Observatory web pages\footnote{http://www.eso.org/sci/facilities/alma/observing/tools/etc/}) we also estimate $SN^{\textrm{peak}}_{24}$, defined as the surface brightness signal-to-noise that may be obtained at $y_{\textrm{peak}}$ using ALMA within an exposure time of 24 hours. We calculate this value assuming continuum observation centred on $\nu = 400$ GHz (within ALMA Band 8), and an antenna configuration giving a beam with FWHM = 0.75 arcsec. In calculating $SN^{\textrm{peak}}_{24}$ each bubble is first convolved with this simulated Gaussian beam to estimate the observable brightness temperature excess at this resolution.  The sensitivity at the continuum band centred on $\nu$ is calculated using the online observing tool, taking the declination (and hence typical atmospheric column density) of each object into account. For those targets with a sky position that cannot be observed using the Southern Hemisphere ALMA facility ($\delta \simeq 30^{\circ}$ and above) we instead use an equatorial location for comparison.

\begin{figure}
\begin{center}
\vspace*{-3mm}
\psfig{figure=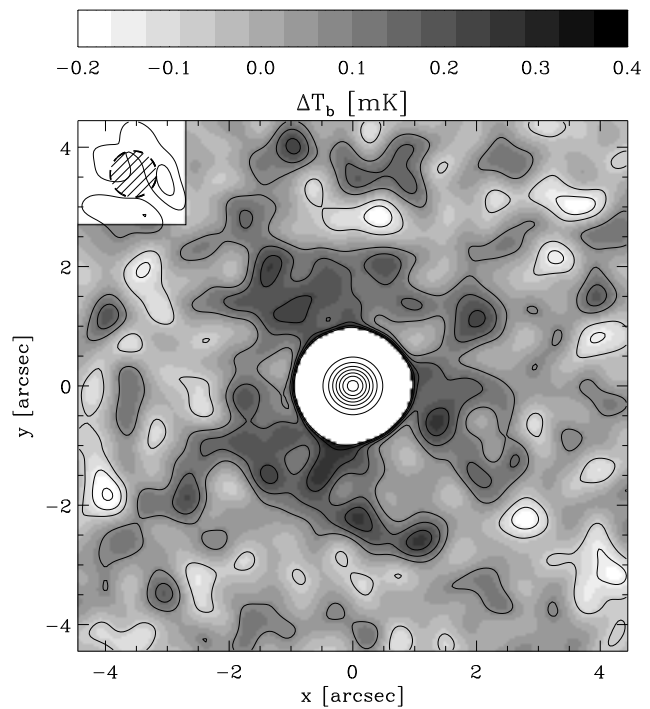,angle=0,width=8.6cm}
\parbox{8.4cm}{\caption{Simulated ALMA excess brightness temperature map at 400 GHz, relative to the CMB, for SMM J222174+0015 (as described in Section \ref{sect:targets}). The simulated exposure time is $5 \times 24$ hours, 
and contours in the greyscale-shaded region are placed at 1$\sigma$ intervals around $\Delta T_{\textrm{b}} = 0$. We remove the greyscale map in the central, SMG dust-contaminated region subject to a threshold of $\Delta T_b \le 0.4$ mK, but contours are here plotted at intervals of 100 mK to illustrate this bright central source. The hatched area  in the upper left illustrates the simulated beam of FWHM $=0.75$ arcsec. \label{fig:shockplot}}}
\end{center}
\end{figure}

To visualize these results we simulate an ALMA image of one of the bubbles visible from the Southern Hemisphere, SMM J222174+0015 \citep{nerietal03,solomonvandenbout05}, at the same $\nu$ and beamsize but for an extended exposure time of $5 \times 24$ hours: this bubble can be seen in Figure \ref{fig:shockplot}. The map also includes contamination due to warm dust emission in the hyper-starburst SMG, for which we assume dust grains at 40 K with emissivity $\epsilon_{\nu} \propto {\nu}^{\beta}$ and spectral index $\beta = 1.5$ (e.g.\ \citealp{blainetal02}). We model the emitting surface brightness distribution as a uniform disc of diameter $2$ kpc in the centre of the bubble, comparable with recent observations of compact hyper-starbursts \citep{caseyetal09,walteretal09,maiolinoetal07}.  Taking the value of $6.3 \pm 1.3$ mJy for the SMM J222174+0015 flux density at 850 $\mu$m \citep{solomonvandenbout05} we estimate a flux density of 9.7 mJy at 400 GHz, and plot contours of the convolved brightness temperature distribution in the central region of Figure \ref{fig:shockplot}.  The brightness temperature greyscale is here removed subject to a simple cut of $\Delta T_b \le 0.4$ mK showing that, thanks to the angular resolution of ALMA, the extended bubble is discernible beyond the strong-but-localized dust emission at its centre.
In practice, a more sophisticated removal of SMG dust contamination will be aided by the possibility of ultra-high resolution mapping of the bright central dust (ALMA can reach FWHM = 0.011 arcsec at 400 GHz), and by the characteristic signature of the tSZ effect on CMB photons: a maximum spectral excess/decrement around 385/144 GHz and null crossover at 218 GHz \citep{carlstrometal02}. Multiple observations will require further telescope time, however.

In Section \ref{sect:sz} it was noted that relativistic corrections to the tSZ are not included in the determination of $y$.  Using equation \eqref{eq:rankine}, the temperature of the gas in the shock boundary is $T_2 = 3 (m_e + m_p) \dot{R}^2_2 / (4 k_{\textrm{B}}) $, which implies a temperature of $T_2 \gtrsim 10^7$K for all the objects listed in Table \ref{tab:targets}.  Calculating the reduction in tSZ intensity using the first and second order relativistic terms described by \citet{challinorlasenby98} yields corrections of $\simeq 3 \%$ for the hyper-starbursts listed.  The effect is therefore negligible for the current model, but relevant for more detailed calculations.

\section{Radiative cooling in the injection bubble}\label{sect:cooling}
The discussion so far has assumed adiabatic bubble expansion, but depends upon losses due to radiative cooling being small within the hyper-starburst age $t$.  At the shock boundary temperatures of $T_2 \gtrsim 10^7$K the gas will be fully ionized but not sufficiently energetic for electron-positron annihilation to contribute to radiative processes: thermal bremsstrahlung and inverse-Compton cooling by CMB photons will dominate radiative losses (this latter being the net cooling effect of the tSZ itself).  The radiative cooling will be greatest at the shock boundary $R_2$, where the combination of plasma temperature $T_2$ and density $\rho_2$ are greatest, and so we calculate timescales for radiative losses in this region to explore the validity of the adiabatic expansion approximation.

We consider first the effect of inverse-Compton scattering of CMB photons. In the relevant non-relativistic limit, the net energy loss rate for a single electron in the shock plasma is given by 
\begin{equation}
-\frac{\dif E}{\dif t} = \frac{16}{3} \sigma_{\textrm{T}} \sigma_{\textrm{SB}} T^4_{\textrm{CMB}}\frac{\expect{v^2_e}}{c^2}
\end{equation}
(see, e.g., \citealp{longair92}) where $\sigma_{\textrm{T}}$ is the Thomson scattering cross-section, $\sigma_{\textrm{SB}}$ is the Stefan-Boltzmann constant, $T_{\textrm{CMB}}$ is the CMB photon temperature at that epoch, and $v_e$ is the velocity of the electron.  This gives an exponential decay law with an inverse-Compton cooling timescale of
\begin{equation}\label{eq:tic}
\tau_{\textrm{IC}} =  \left(3 m_e c^2 \right) / \left[ 32 \sigma_{\textrm{T}} \sigma_{\textrm{SB}} T^4_{\textrm{CMB}}(z) \right]
\end{equation}
at the epoch at which the shock is observed. Values of $\tau_{\textrm{IC}}$ for the selection of targets discussed in Section \ref{sect:targets} can also be seen in Table \ref{tab:targets}, and in all cases can be seen to be large in comparison to the typical hyper-starburst lifetime of $\sim$$10^7$ years.

Cooling due to thermal bremsstrahlung near the shock boundary $R_2$ may be more significant.  A thermal hydrogen electron in the shock boundary plasma will lose energy at a rate
\begin{equation}\label{eq:brem}
-\frac{\dif E}{\dif t} = 1.435 \times 10^{-40} n_p \bar{g} \sqrt{T_2} ,
\end{equation}
where $n_p$ is the volume number density of protons in the plasma and $\bar{g} \simeq 1.2$ is taken as a typical Gaunt factor \citep{longair92}. In Table \ref{tab:targets} we use this relation to give timescales $\tau_{\textrm{B}}$ over which the shock boundary energy would be reduced to 37\% of its instantaneous adiabatic value by bremsstrahlung alone, as an indicator of the importance of radiative cooling.  These timescales are generally longer than the typical lifetimes of hyper-starbursts, but not always: bremsstrahlung will clearly dominate over inverse-Compton losses, and at higher redshift (where $n_p$ is greater) will limit the epoch of pure adiabatic evolution to within $t \lesssim 10^7$ years, comparable with the duration of hyper-starburst activity. However, these radiative losses are not total: the adiabatic model can be viewed as giving an approximate upper bound on the expected range of $y_{\textrm{peak}}$. This increases the importance of bubble age in the tSZ signal beyond the otherwise weak $y \propto t^{-1/5}$, and might be used to provide lower bounds on $t$ should a tSZ bubble not be detected.

\section{Discussion \& Conclusions}\label{sect:conc}
We have presented an adiabatic model of wind-driven bubble shockwaves around hyper-starburst galaxies and made predictions for the strength of the tSZ effect due to such objects.  Assuming this model to be a reasonable approximation we have shown that such bubbles are detectable using ALMA, albeit requiring integration times of multiple days (in two or, preferably, three bands). We have shown that bremsstrahlung will dominate the radiative cooling of the shocked gas, and found that this will cause the evolution of the shock to become significantly non-adiabatic after a time comparable to the typical total duration of hyper-starbursts themselves. 

In order to construct the model a number of simplifications have been made. Among these, the assumption of spherical symmetry and the homogeneous density $\rho_d$ are least likely to be valid.  The former assumption may be thrown into question for systems with active galactic nuclei, as are often seen for hyper-starbursts, or with star formation occurring in an extended disc rather than a central core: \citet{dallavecchiaschaye08} found a bi-polar pressure distribution when modelling outflows in massive disc-like formation, once winds were fully coupled to the gas dynamics of the system. A bi-polar outflow would alter the pressure distribution of the shocked gas and might reduce the strength of the tSZ effect along certain lines of sight.

However, the star typical star formation rates in the massive systems simulated by these authors were small when compared to the figures in Table \ref{tab:targets}, and it might be expected that the greater mechanical energy input from hyper-starbursts will alter the wind-IGM interaction significantly. \citet{dallavecchiaschaye08} found interesting qualitative differences in the behaviour of outflows for their simulated disc systems of total halo mass 10$^{10}h^{-1}$M$_{\odot}$ and 10$^{12}h^{-1}$M$_{\odot}$ that produced early star formation rates of $\simeq 0.05$~M$_{\odot} \textrm{yr}^{-1}$ and $\simeq 10 $~M$_{\odot} \textrm{yr}^{-1}$ respectively.  An increase of a further two orders of magnitude in star formation rate may cause sufficient increases in wind ram pressure as to qualitatively change the outflow behaviour once more. In a related study \citet{chatterjeekosowsky07} and \citet{chatterjeeetal08} found that quasar-driven bubbles in simulated halos provided a reasonable match to a simple spherical model, although again at lower energies and typically after longer timescales than we consider here.  Unfortunately there is insufficient knowledge and modelling of the wind-interstellar medium interaction in these powerful systems to do more than speculate.  The typical morphologies and environments of the star-forming gas regions prior to the onset of strong winds are also poorly known for these hyper-starbursts, which further complicates accurate simulation.  More detailed theoretical modelling combined with future precision observations of the tSZ might help shed light on exactly these questions.

Another simplification, the assumption of a homogeneous IGM, may misrepresent the evolution of the bubble with time: for a more realistic density profile the bubble will initially expand more slowly through the dense gas at the halo centre, and then more rapidly at later times as $\rho_d(R_2)$ decreases. This is an area in which the modelling could be improved, particularly if aided by more detailed knowledge of the gas density around hyper-starbursts.  A related simplification is the model's neglect of the gravitational deceleration of the shockwave due to the host halo potential.  This effect will generally represent a small correction to the dynamics of adiabatic expansion and radiative losses, which can be seen by comparing the galaxy wind velocities to the escape velocities of most galactic haloes (entrained clouds, however, might impart a velocity kick that prevents their escape). These effects should be considered in more accurate modelling, perhaps using a simplified model of a galaxy halo potential and mass entrainment (e.g.\ \citealp*{murrayetal05}).

Important uncertainties also arise due to the assumed values of the starburst age $t$, and the fiducial parameters $v_{\mw}$ and $\dot{M}_*$ (or equivalently $f_{\textrm{w}}$).  It should be emphasized that the fiducial predictions may be adapted as appropriate using the simple scaling relations presented, which may be necessary as we gain a clearer picture of hyper-starburst physics. As an example, using $y \propto f_{\mw}^{3/5}$ from equation \eqref{eq:ypeak} and adopting the value $f_{\mw} \simeq 0.4$ of \citet{dallavecchiaschaye08} reduces the tSZ signals listed in Table \ref{tab:targets} by $\simeq 20\%$. However, a drastic reduction in $f_{\mw}$ (appropriate if the kinetic energy from hyper-starburst supernovae is mostly lost radiatively) might yet reduce $y$ to the cosmological confusion limit calculated by \citet{roncarellietal07}.
Conversely, observations of the tSZ around hyper-starbursts could in this way, even if not detected, provide interesting constraints on $f_{\textrm{w}}$ and thus supernova wind feedback in these objects, which may contribute much of the total star-formation at high redshift \citep{blainetal02}.  They might also constrain the feedback energy injected by the active galactic nuclei commonly hosted in hyper-starburst systems (e.g.\ \citealp{chatterjeeetal08}), although more careful modelling will be required to distinguish the two signals.


To conclude, wind-driven bubbles around hyper-starburst objects may be detectable via the tSZ in the near future using submillimetre instruments such as ALMA or its successors. Despite many simplifying assumptions in the model, the enormous mechanical energy of the wind-IGM interaction will ensure that a system approximating the simple, self-similar blastwave must persist while radiative losses are small.  According to such a model, such objects as those listed in Table \ref{tab:targets} (with estimated $y_{\textrm{peak}} > 2 \times 10^{-5}$) might be securely detected with ALMA within days. The number of known hyper-starbursts in the Southern Hemisphere must be expected to increase with time, and one of these may present an excellent opportunity for a first bubble detection with ALMA.

\section{Acknowledgements}
The authors would like to thank Richard Bielby, Arthur Kosowsky, Vivienne Wild, and the anonymous referee for useful comments and suggestions. Barnaby Rowe has been  supported in part by the Dark Universe through Extragalactic Lensing (DUEL) European Union FP6 Research Training Network (MRTN-CT-2006-036133).

\bsp

\bibliographystyle{mn2e}

\bibliography{btprmnras_letter}

\end{document}